\documentclass[12pt]{article}
\usepackage{epsfig}
\input epsf.sty
\newcommand{\la}[1]{\label{#1}}
 
\newcommand{\be}{\begin{equation}}
\newcommand{\ee}{\end{equation}}
\newcommand{\ba}{\begin{eqnarray}}
\newcommand{\ea}{\end{eqnarray}}
\newcommand{\bi}{\begin{itemize}}
\newcommand{\ei}{\end{itemize}}
\newcommand{\rmi}[1]{{\mbox{\scriptsize #1}}}
\newcommand{\nr}[1]{(\ref{#1})}

\renewcommand{\vec}[1]{{\bf #1}}

\newcommand{\eq}{Eq.~}
\newcommand{\eqs}{Eqs.~}
\newcommand{\fig}{Fig.~}
\newcommand{\figs}{Figs.~}

\def\lsi{\raise0.3ex\hbox{$<$\kern-0.75em\raise-1.1ex\hbox{$\sim$}}}
\def\gsi{\raise0.3ex\hbox{$>$\kern-0.75em\raise-1.1ex\hbox{$\sim$}}}

\makeatletter \@addtoreset{equation}{section} \makeatother
\renewcommand{\theequation}{\arabic{section}.\arabic{equation}}
 
\begin{document}
 
\begin{titlepage}
\begin{flushright}
CERN-TH/2000-080\\
NORDITA-2000/29HE\\
hep-lat/0003020
\end{flushright}
\begin{centering}
\vfill
 
{\bf O(2) SYMMETRY BREAKING VS.\ VORTEX LOOP PERCOLATION} 

\vspace{0.8cm}

K.~Kajantie$^{\rm a,}$\footnote{keijo.kajantie@helsinki.fi},
M.~Laine$^{\rm b,a,}$\footnote{mikko.laine@cern.ch},
T.~Neuhaus$^{\rm c,d,}$\footnote{neuhaus@physik.rwth-aachen.de}, 
A.~Rajantie$^{\rm e,}$\footnote{a.k.rajantie@sussex.ac.uk},
K.~Rummukainen$^{\rm f,g,}$\footnote{kari@nordita.dk} \\

\vspace{0.3cm}
{\em $^{\rm a}$Dept.\ of Physics, P.O.Box 9, 
FIN-00014 Univ.\ of Helsinki, Finland\\ }
\vspace{0.3cm}
{\em $^{\rm b}$Theory Division, CERN, CH-1211 Geneva 23,
Switzerland\\ }
\vspace{0.3cm}
{\em $^{\rm c}$Institut f\"ur Theoretische Physik E, RWTH Aachen, FRG\\ }
\vspace{0.3cm}
{\em $^{\rm d}$ZiF, Univ.\ Bielefeld, D-33615 Bielefeld, FRG\\ }
\vspace{0.3cm}
{\em $^{\rm e}$Centre for Theor. Physics, Univ. of Sussex, 
Brighton BN1 9QH, UK\\ }
\vspace{0.3cm}
{\em $^{\rm f}$NORDITA, Blegdamsvej 17,
DK-2100 Copenhagen \O, Denmark\\ }
\vspace{0.3cm}
{\em $^{\rm g}$Helsinki Inst.\ of Physics,
P.O.Box 9, FIN-00014 Univ.\ of Helsinki, Finland\\ }

\vspace{0.7cm}
{\bf Abstract}
 
\end{centering}
 
\vspace{0.3cm}\noindent
We study with lattice Monte Carlo simulations the relation 
of global O(2) symmetry breaking in three dimensions to the 
properties of a geometrically defined vortex loop network. We 
find that different definitions of constructing a network lead 
to different results even in the thermodynamic limit, and that with 
typical definitions the percolation transition does not coincide 
with the thermodynamic phase transition. These results show that 
geometrically defined percolation observables need not display 
universal properties related to the critical behaviour of the 
system, and do not in general survive in the field theory limit.
\vfill
\noindent
 
\vspace*{1cm}
 
\noindent
CERN-TH/2000-080\\
NORDITA-2000/29HE\\
March 2000
 
\vfill

\end{titlepage}
 
\section{Introduction}

In a classic work~\cite{bmk}, Banks, Kogut and Myerson showed that
the partition function of a particular three-dimensional (3d) spin
model with global O(2) symmetry, called the Villain model, can 
equivalently be represented as a partition function of a dual theory
in which the spin configurations are integer-valued and sourceless
and can therefore be represented as configurations of closed loops.
Moreover, they suggested that at the transition point where the 
global O(2) symmetry gets restored, infinitely long loops become
important. The loops of the dual theory can be identified 
with vortex lines of the original theory, and therefore this
result suggests that also in other theories that contain
topological defects, they might play an important role in
determining the properties of the phase transition.
As a consequence, many problems in condensed matter physics 
and cosmology could be simplified by concentrating 
only on the vortex loop degrees of freedom 
(see, e.g., \cite{Williams:1999} and references therein). 

In many realistic
cases, this intuitive picture can be realized
in the precise sense that the vortex tension $T$ defined 
as the energy per unit length of an isolated vortex vanishes 
at the transition point~\cite{thooft,cgpt,tension}.
Near the transition point, vortex loops behave as
world lines of particles in 2+1d, and following 
the Villain case~\cite{kks_dual} one can then
conjecture that they could be described with 
an effective scalar field of 
mass $T$~\cite{kks_dual,kovner}. However, it is rather difficult to 
measure the tension $T$ in 
practice~\cite{tension,Kovacs:2000sy,manyvortex}, 
which limits the practical uses of this approach.

Therefore, one may seek for alternative ways of studying
the vortex degrees of freedom. A seemingly natural approach
is to decompose every spin configuration generated in a lattice
Monte Carlo simulation into a number of
closed vortex loops. The hope is then that the transition 
could be identified
with a non-zero probability of finding vortex loops
that extend through the whole system, 
a phenomenon which is often called percolation.

Percolation has been used to study phase transitions in various 
different theories. Examples of this in condensed matter physics include 
the 3d XY (or Heisenberg) model~\cite{pochinsky,hu,ns}, and in cosmology 
the corresponding global O(2) field theory~\cite{Antunes},
both assumed to represent some features of the gauged Ginzburg-Landau 
theory where the loops would be interpreted as 
Nielsen-Olesen vortices. In particle physics, a similar approach 
has been used to give a physical interpretation for the confinement 
phase transition in non-Abelian gauge theories~\cite{Engelhardt:2000fd}
and to define an order parameter in the crossover regime
of the electroweak theory~\cite{cgis2}.

Given the large range of physical applications for the percolation
picture, it is important to check whether it really works.
Experience from the Ising model where a different kind of 
percolation may occur, related to clusters instead of vortex 
loops, has shown~\cite{ref:muller} that one has to be quite 
careful with this interpretation (for a recent discussion, 
see~\cite{fs}). In the present paper we study with high precision 
lattice Monte Carlo simulations the geometric properties of a properly 
defined vortex loop network near the thermodynamic phase transition 
in the XY model and show that in all of the cases considered, the 
percolation point fails to coincide with the thermodynamic singularity.
Nevertheless, a description in terms of vortex degrees of freedom 
can still be useful in many other contexts, if used with caution.

\section{Model} 
\la{model}

The 3d XY model is defined by 
\be
Z_\rmi{XY} =  \int_{-\pi}^{\pi} \!\Pi_\vec{x}\, d\theta_\vec{x}  
\exp\Bigl(
\beta \sum_\vec{x}\sum_{i=1}^3 
\cos ( \theta_\vec{x+i}-\theta_\vec{x})\Bigr). 
\la{XY}
\ee
It belongs to the same universality class as the Villain 
model where the original consideration involving vortex loop 
degrees of freedom was carried out~\cite{bmk}. However, it is 
calculationally simpler, its critical properties are known very 
accurately, and it is more directly related to continuum field 
theories such as the Ginzburg-Landau model.

Let us recall that the only parameter point of the XY model 
corresponding to a continuum field theory is the 2nd order transition 
point at $\beta=\beta_c$. Thus for obtaining results
free from lattice artifacts, one should focus on the vicinity
of~$\beta_c$. {}From the statistical mechanics point of view, 
this is the region where physical observables show universal 
critical behaviour.

\section{Method}
\la{method}

In order to analyse accurately the universal behaviour close to $\beta_c$, 
we shall employ the finite size scaling method~\cite{d8}. We recall that the
idea of the method is to first locate 
the infinite-volume critical point of the system. One then measures
finite-volume values for various observables at that point, and
determines from them the critical exponents. 

We will here address particularly two questions:
\bi
\item[1.]
Does the infinite-volume extrapolation of the percolation critical point
coincide with the thermodynamic critical point? 
\item[2.]
Sitting at the thermodynamic critical point, do percolation related
observables show critical behaviour?  If so, do the critical exponents
match any of the known thermodynamic ones?
\ei

As a benchmark with which to compare, let us recall the known thermodynamic 
properties of scalar models with global
O(2) symmetry. In the XY case, the infinite
volume critical point is at~\cite{bal}
\be
\beta_c=0.454165(4). \la{betac}
\ee
Let us denote 
\be
E = -\sum_\vec{x}\sum_{i=1}^3 
\cos ( \theta_\vec{x+i}-\theta_\vec{x}), \quad
M = \sum_\vec{x} \cos (\theta_\vec{x}), \la{EM}
\ee
where we have assumed the symmetry breaking direction
to be $\theta=0$.
For any operator $O_\rmi{i}$, one can define the 
related susceptibility, 
\be
C_\rmi{i} = N^3 \Bigl\langle\Bigl( 
O_\rmi{i} - \langle O_\rmi{i} \rangle \Bigr)^2\Bigr\rangle, 
\la{susdef}
\ee
where $N^3$ is the volume of the system. 
For energy ($E$) and magnetization ($M$) 
like observables at $\beta_c$, the susceptibilities diverge as 
\be
C_E \propto N^{\alpha/\nu}, \quad
C_M \propto N^{\gamma/\nu}. \la{gamma}
\ee
Here $\nu$ is the 
correlation length critical exponent. 
Magnetization itself scales at the critical point 
according to 
\be
\langle |M|\rangle \propto N^{-\beta/\nu}. 
\ee
Finally, corrections
to scaling are determined by an universal exponent $\omega$; 
for instance, the finite volume apparent critical point
determined from $C$ is assumed to behave as 
\be
\beta^{N}_c -\beta^{N=\infty}_c = c_1 \frac{1}{N^{1/\nu}} + 
c_2 \frac{1}{N^{1/\nu+\omega}} + ...\, . \la{fss}
\ee
The known results for the exponents appearing here are~\cite{bal,ZJ,gh}:
\smallskip

\be
\begin{minipage}[t]{10cm}
\centerline{
\begin{tabular}{cccccccc}
\hline
$\alpha$ & $\beta$ & $\gamma$ & $\nu$ & $\omega$ 
& $\alpha/\nu$ & $\beta/\nu$ & $\gamma/\nu$ \\
\hline
-0.01 & 0.35 & 1.32 & 0.67 & 0.79 & -0.02 & 0.52 & 1.97 \\
\hline
\end{tabular}}
\end{minipage}
\la{table}
\ee

\section{Observables}
\la{sec:obs}

\begin{figure}

\vspace*{0.5cm}

\begin{center}
\epsfig{file=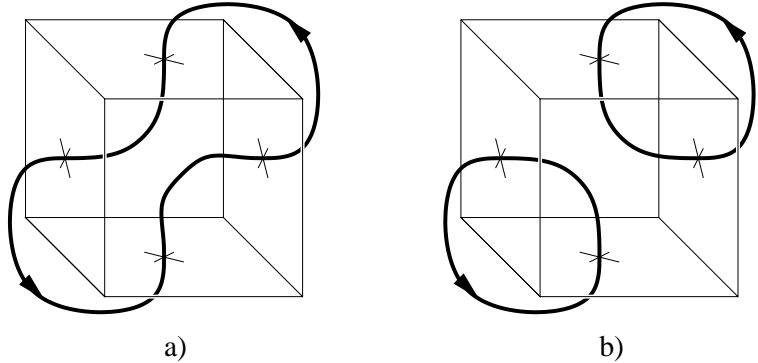,width=10cm}
\end{center}
\caption{
\label{fig:connect}
If two vortices pass through one cell, the vortex tracing algorithm must
decide how to connect them, and this leads to an ambiguity
in the length distribution.}
\end{figure}

In order to locate the vortices in a given field configuration,
we first define 
\be
Y_{i,\vec{x}}=[\theta_\vec{x+i}-\theta_\vec{x}]_\pi,
\ee
where $[...]_\pi$ means $(...)$ mod $2\pi$ 
with a result in the interval $(-\pi,\pi]$. 
The winding number of a plaquette is then defined as
\be
n_{ij,\vec{x}}=\frac{1}{2\pi}\left(Y_{i,\vec{x}}+Y_{j,\vec{x+i}}
-Y_{i,\vec{x+j}}-Y_{j,\vec{x}}
\right).
\label{ndef}
\ee
The simplest vortex-related quantity is now~\cite{ksw}
\be
\langle O_\rmi{tot}\rangle = \langle |n_{ij,\vec{x}}| \rangle. \la{Otot}
\ee
It has the interpretation of the total vortex density, 
independent of their length. It is a perfectly 
well defined local observable, and any singularities
in its finite volume behaviour are expected to be related to the 
thermodynamic singularities of the system.

However, in the continuum limit ($t = \beta-\beta_c \rightarrow 0$)
$O_\rmi{tot}$ smoothly approaches a constant value $\approx 0.160$,
and the interpretation of $O_\rmi{tot}$ as a vortex density becomes
problematic: if we rescale the length scale by the physical
correlation length $\xi\sim t^{-\nu}$, the vortex density diverges
as $t^{-2\nu}$.  {}From the continuum physics point of view, the density
of vortex lines increases as we increase the resolution at which we
look at the system, and finally we are measuring only ultraviolet
fluctuations. This implies that quantities like the number of vortex 
loops and the length of individual loops are not well defined.
The same problem arises also in more general continuum theories 
(see \cite{vortex} and references therein). 

These problems might be circumvented if we only monitor
sufficiently long loops. Thus, we next connect single plaquettes 
with $n_{ij,\vec{x}}\neq 0$ to a macroscopic closed
vortex loop. This steps 
involves a lot of ambiguity, see \fig\ref{fig:connect}. 
We will employ two different ``connectivity'' definitions here:
\bi
\item[I.] In the {\em maximal} definition, 
in each ambiguous case we make the unique choice so
that the separate vortex loops are joined together,
see \fig\ref{fig:connect}.a.
\item[II.] In the {\em stochastic} definition, in each
ambiguous case the choice is made by throwing a dice. This is 
the procedure most commonly followed in the literature.
\ei

We can thus extract from each lattice configuration a set of
``vortex loops'', consisting of a number of plaquettes which 
have been bound together by one of the connectivity definitions
above. From this set of loops, we measure (using always
cubic volumes $N^3$) the following observables:

{\bf 1.} ``Closed loop order parameter'', $O_\rmi{closed}$. 
For a given configuration, $O_\rmi{closed}=1$
if there is a vortex which 
forms a closed {\em non-contractible\,} loop around the lattice with
periodic boundary conditions (in fact, there have to be at least
two such loops, because the total topological charge is zero),
and $O_\rmi{closed}=0$ otherwise. 
We observe that
$\langle O_\rmi{closed}\rangle$ 
approaches a step-like function at $N\to\infty$, 
being non-zero in the symmetric phase of the XY model.
We can also consider a variant of this observable in which we count
all vortices that extend over 
a distance $\ge N$ in any direction ($O_{xyz}$) or in
some particular direction ($O_{x}$).

{\bf 2.} ``Volume'', $O_\rmi{vol}$. We fit each vortex loop to the smallest
possible rectangular box $n_1\times n_2\times n_3$. Let $O_\rmi{vol}$
be the volume of the largest box compared with the total, 
$O_\rmi{vol} = n_1n_2n_3/N^3$. We again
observe that $\langle O_\rmi{vol}\rangle$ approaches a step function
in the infinite-volume limit.

{\bf 3.} ``Mass'', $O_\rmi{mass}$. 
For each configuration, we count the number 
of plaquettes visited by the longest loop $L_0$, and normalise by the 
volume: 
$O_\rmi{mass} = (1/N^3) \sum_{\rmi{plaq.}\in L_0}$. 
We observe a vanishing result in the broken phase for $N\to\infty$, 
a non-vanishing in the symmetric phase, but 
a continuous behaviour at $\beta_c$. 
Thus $\langle O_\rmi{mass}\rangle$
behaves like the magnetization $M$ but with an ``inverted''
$\beta$-axis. 

{\bf 4.} ``Line tension'', $T_L$.
If in the idealised case of non-interacting loops, 
the tension, i.e.~the mass per unit length, of
a vortex is $T$, then the loop length
distribution $\rho(L)$ has the form~\cite{ns}
\be
\rho(L)\sim L^{-5/2}\exp(-LT).
\ee
In the interacting case, the line tension $T_L$ is defined by 
fitting the loop distribution $\rho(L)$ to a similar function,
\be
\rho(L)\sim L^{-\alpha}\exp(-LT_L),
\la{loopdist}
\ee
where $\alpha$ and $T_L$ are free parameters. However, 
in the interacting case $T_L$ should not be confused with the tension~$T$ 
given by the free energy per unit length of an isolated 
vortex, and the two do not in general coincide. We observe that the 
line tension behaves similarly to the real (not the inverted) 
magnetization $M$.

For any observable $O_\rmi{i}$, 
$\mbox{i}=\{\mbox{closed}, xyz, x, \mbox{vol}, \mbox{mass}\}$, 
we define the percolation point $\beta_\rmi{i}$ as
the point where $\langle O_\rmi{i}\rangle$ vanishes when $\beta$ is increased.
Because of the similarities in the definitions of the observables, we
expect there to be certain inequalities. Even in a finite system
and in any configuration, 
the observables satisfy $O_{xyz}\ge O_\rmi{closed}$,
$O_{xyz}\ge O_x$, as well as
$O_\rmi{vol}\ge O_\rmi{mass}/3$ because each lattice cube has three
plaquettes. The lattice symmetry also implies
that $\langle O_x\rangle\ge \langle O_{xyz}\rangle/3$, 
because on the average at least every third long loop 
has to extend into the $x$-direction. These relations imply
\be
\beta_\rmi{mass}\le \beta_\rmi{vol},
\quad
\beta_\rmi{closed}\le\beta_{xyz}=\beta_x.
\la{inequ0}
\ee
It also seems very plausible that in the infinite-volume limit,
almost all configurations have either $O_\rmi{vol}=0$ or $1$
and consequently $O_{xyz}\ge O_\rmi{vol}$.
We may also conjecture that in the infinite volume limit
$\langle O_\rmi{vol}\rangle=\langle O_{xyz}\rangle$, 
and that $\langle O_\rmi{mass}\rangle > 0$ implies
$\langle O_\rmi{closed}\rangle > 0$ because long volume
filling loops have a finite probability to close. 
Therefore, we may expect that
\be
\beta_\rmi{mass}\le
\beta_\rmi{closed}\le \beta_\rmi{vol}=\beta_{xyz}=\beta_x,
\la{inequ}
\ee
and we cannot rule out the possibility that all these points are equal.
However, in our actual analysis we only rely on \eq\nr{inequ0}, and use
\eq\nr{inequ} just as a guidance. 

\section{Results}
\la{results}

We have carried out simulations using both Swendsen-Wang and 
Wolff cluster algorithms. We use volumes in the range $4^3...200^3$.
Individual runs have up to $\sim 10^6$ measurements at smaller 
volumes, $\sim 10^4$ at the largest ones. 

\subsection{Calibration} 

As a calibration of our accuracy we first briefly discuss the
thermodynamic critical properties of the XY model, and show that
we can reproduce the known results with good precision. 

As an example, we show
$C_\rmi{tot}$ in \fig\ref{fig:Otot}(a) (cf.\ \eqs\nr{susdef}, \nr{Otot}). 
The critical point, defined as the point where $C_\rmi{tot}$  
obtains its maximum, is fit
at relatively small volumes $4^3...48^3$
according to \eq\nr{fss}, with exponents fixed
according to \eq\nr{table}. We obtain $\beta_c=0.45430(27)$,
$\chi^2/$dof $=0.19$, in agreement with \eq\nr{betac}.
We can also leave both $\nu$ and $\omega$ as free parameters in the fit,
but then the errors become quite large: $\nu=0.83(15)$, $\omega=1.16(42)$,
$\beta_c=0.45389(62)$, $\chi^2/$dof $=0.16$. 
However, they are consistent with the
values in Eqs.~\nr{betac}, \nr{table}.
As another example, we sit precisely at the $\beta_c$ given 
in \eq\nr{betac}, and measure the finite volume behaviour of
magnetization with volumes $20^3...200^3$. We find the exponent
$\beta/\nu=0.5192(15)$, in agreement with \eq\nr{table}.
We conclude that our lattice sizes and statistics are sufficient 
to resolve the physical critical properties of the XY model. 

\subsection{Volume order parameter}

We now move to our actual topic: percolation related
observables. As an illustration of a behaviour similar to 
but not identical with that in \fig\ref{fig:betap}(a), 
we first discuss the volume order parameter 
$O_\rmi{vol}$, determine its percolation 
critical point $\beta_\rmi{vol}$, 
and compare it with $\beta_c$ in \eq\nr{betac}.

\begin{figure}[tb]

\centerline{(a)\begin{minipage}[c]{13.2cm}
    \psfig{file=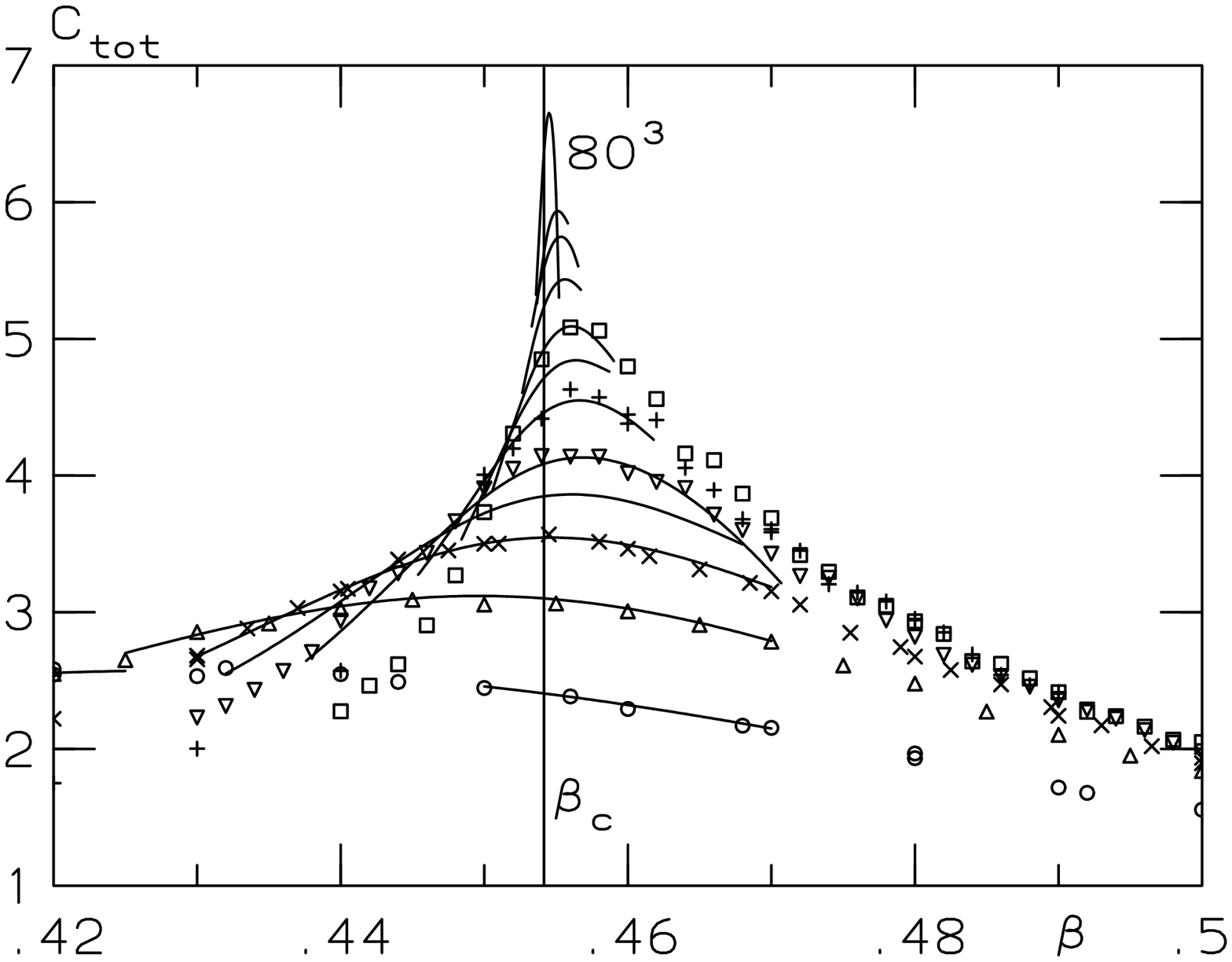,angle=270,width=6.5cm}
    \psfig{file=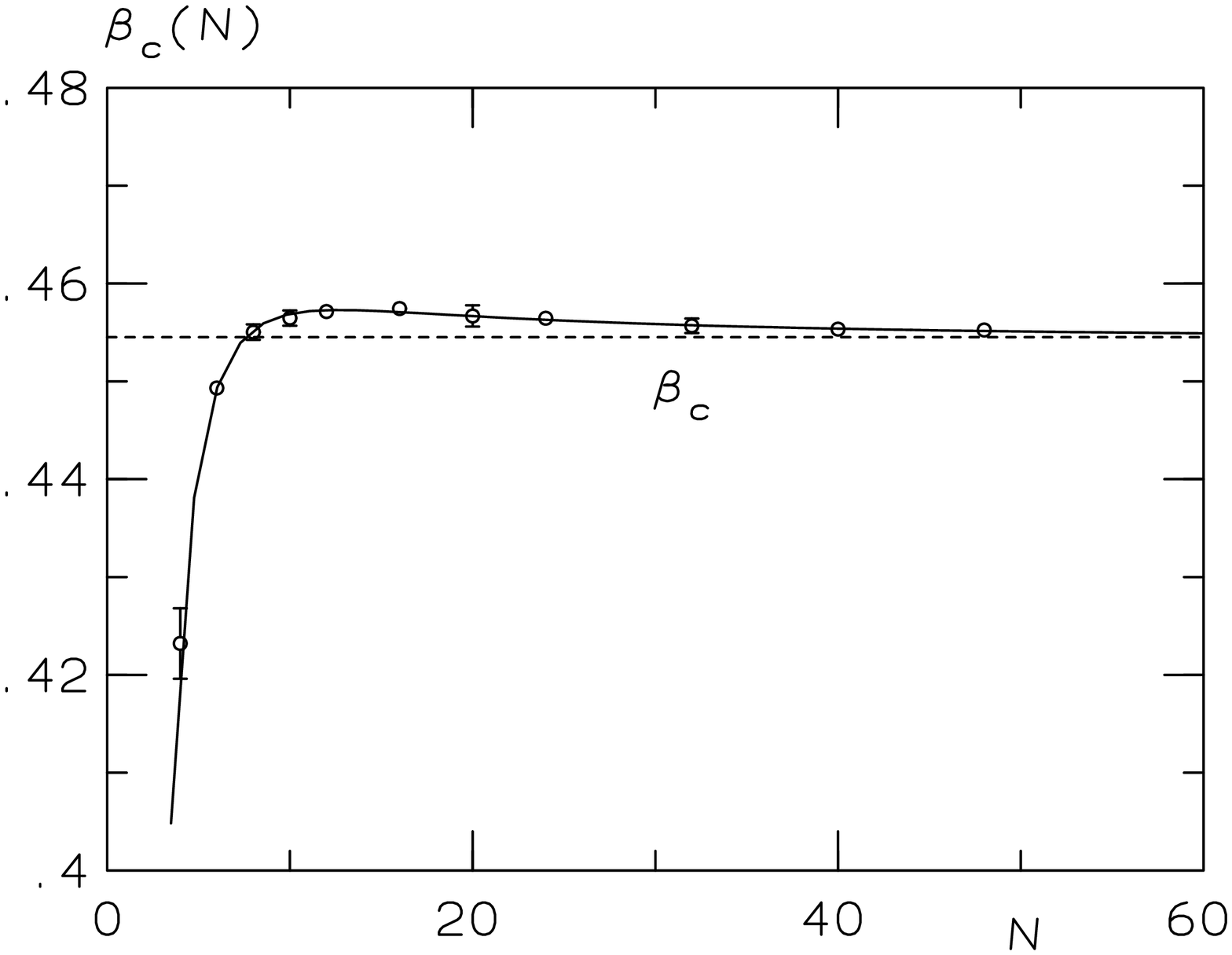,angle=270,width=6.5cm}
    \end{minipage}}

\centerline{(b)\begin{minipage}[c]{13.2cm}
    \psfig{file=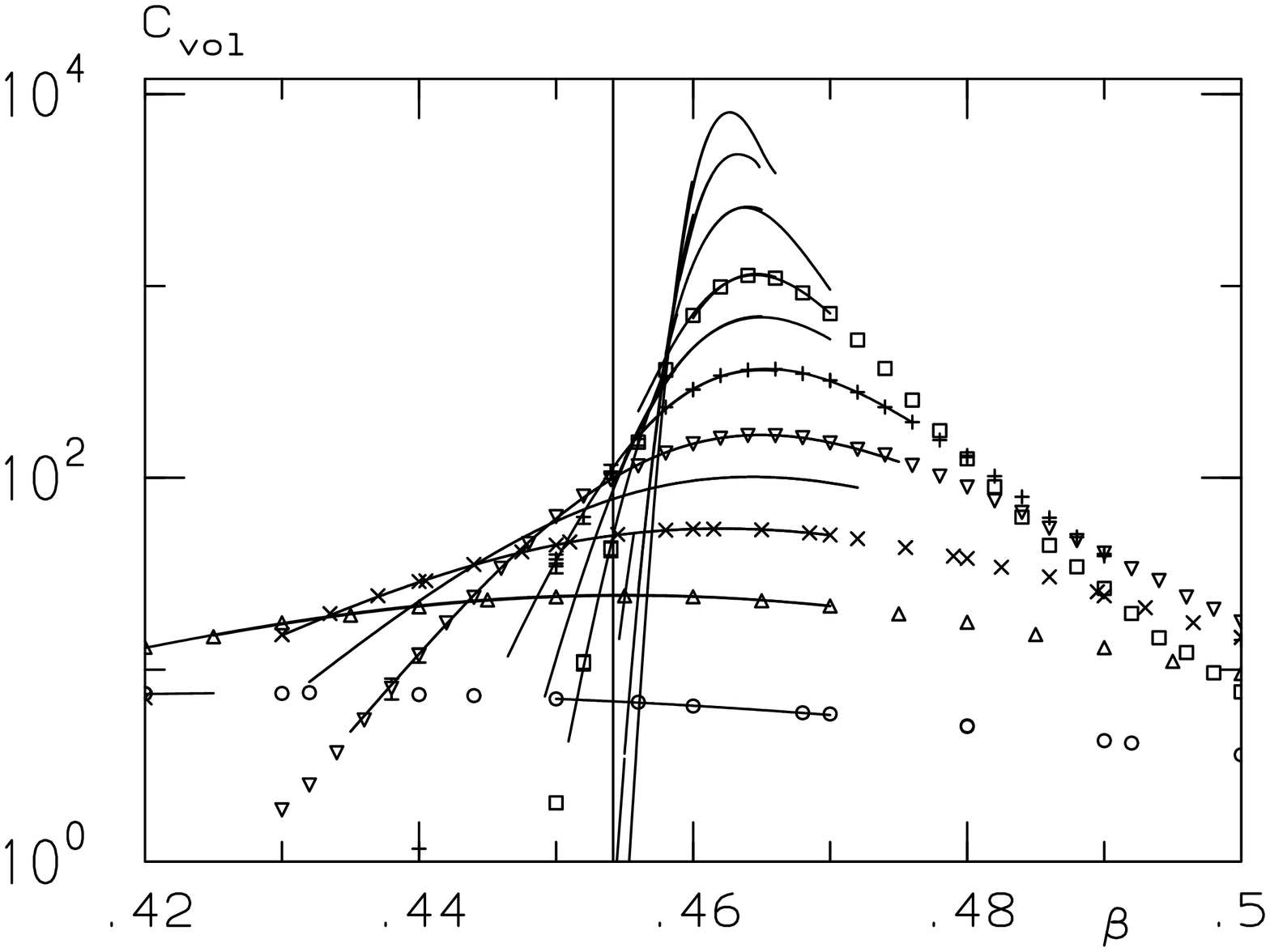,angle=270,width=6.5cm}
    \psfig{file=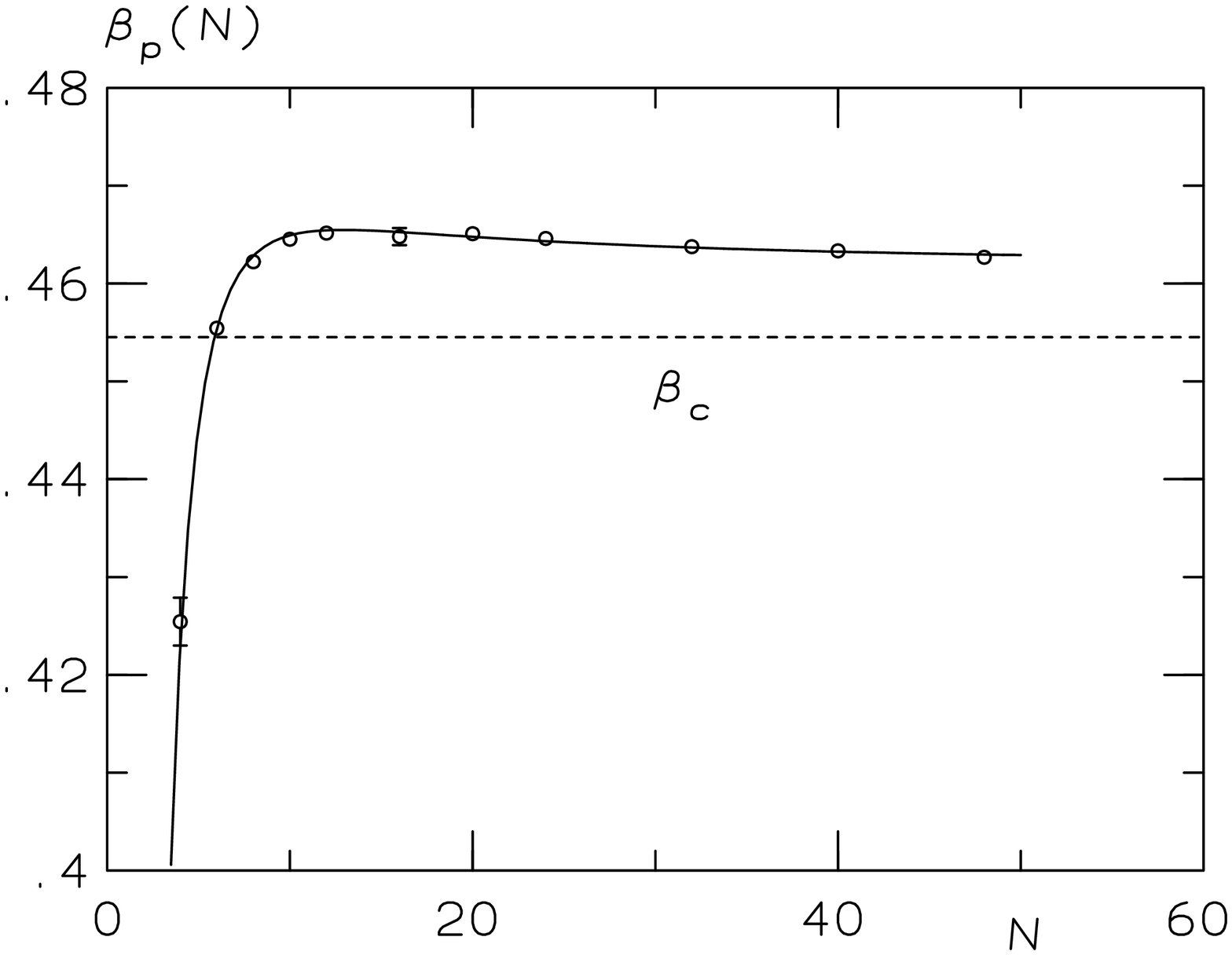,angle=270,width=6.5cm}
    \end{minipage}}

\centerline{(c)\begin{minipage}[c]{13.2cm}
    \psfig{file=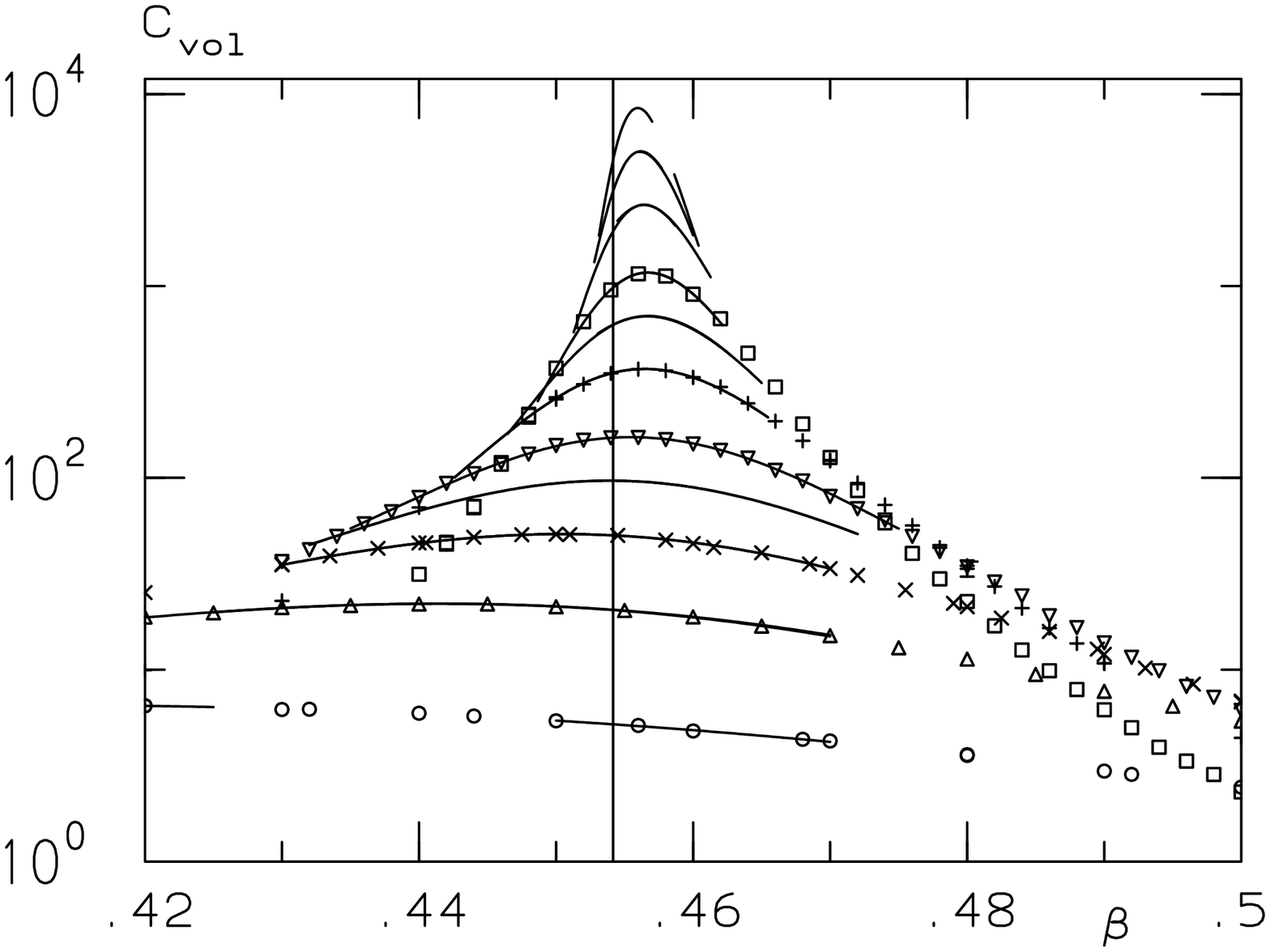,angle=270,width=6.5cm}
    \psfig{file=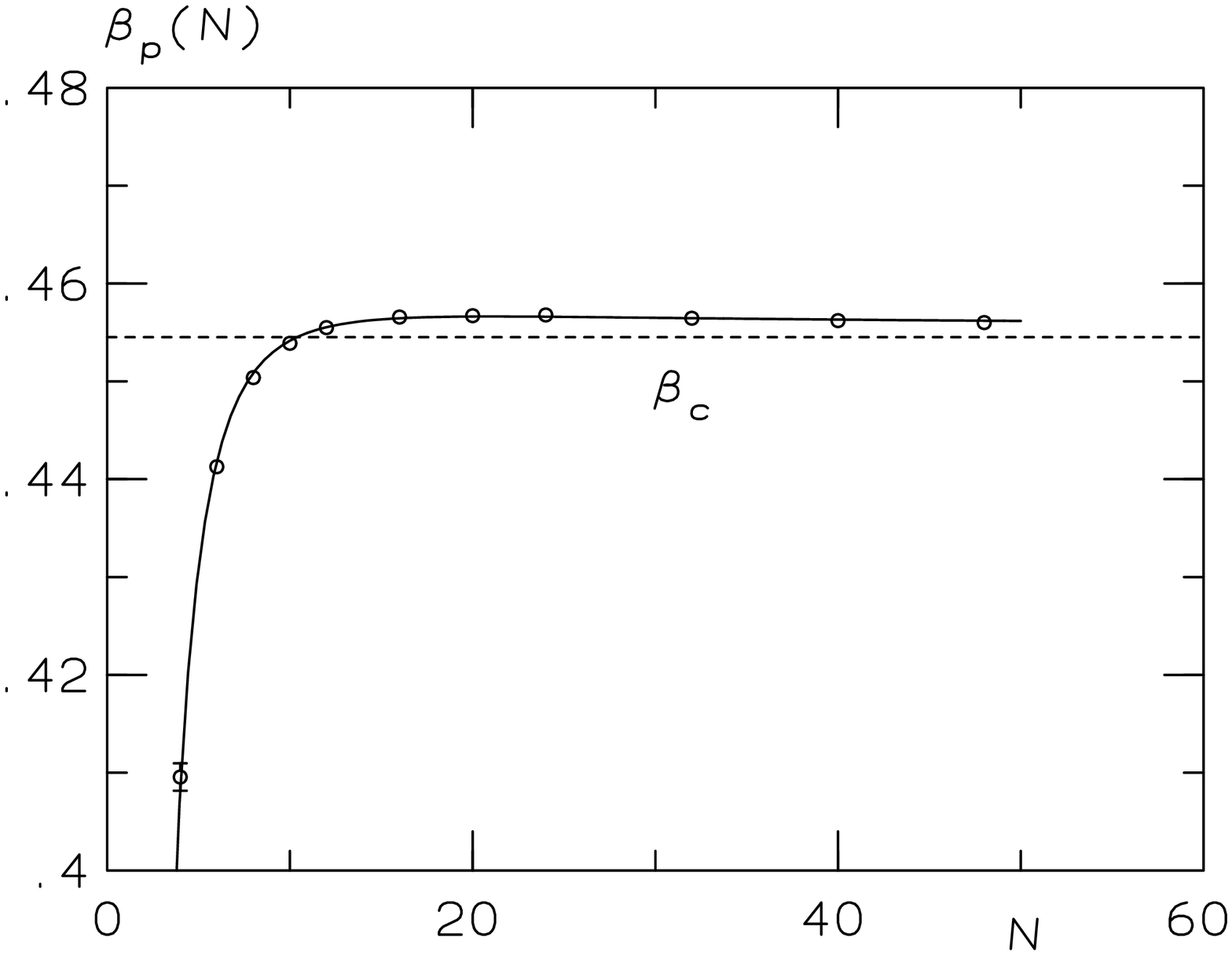,angle=270,width=6.5cm}
    \end{minipage}}

\caption[a]{
Left: different susceptibilities, \eq\nr{susdef},
as a function of $\beta$;  
(a) $C_\rmi{tot}$, (b) $C_\rmi{vol}$ with 
the maximal connectivity definition, (c)
$C_\rmi{vol}$ with the stochastic connectivity definition.
Right: the location of the apparent critical point determined from
the maximum of $C$. The lines indicate the location of $\beta_c$.}  
\la{fig:Otot}\la{fig:betap}
\end{figure}

In \figs\ref{fig:betap}(b),(c)
we show the susceptibility $C_\rmi{vol}$, 
as well as its maximum position, 
measured with both types of connectivity
definitions. We make an infinite volume fit according to \eq\nr{fss}, 
with exponents fixed as above, and another one with 
two free exponents. With the maximal
connectivity definition we obtain 
\ba
\beta_\rmi{vol} & = & 0.46184(14),\quad \nu=0.67, \quad \omega=0.79, \quad
\chi^2/\mbox{dof} = 2.35, \la{mvol1} \\
\beta_\rmi{vol} & = & 0.46074(48),\quad \nu=0.66(3), \quad \omega=0.12(3), 
\quad \chi^2/\mbox{dof} = 0.64, \la{mvol}
\ea
and with the stochastic definition 
\ba
\beta_\rmi{vol} & = & 0.45565(13),\quad \nu=0.67, \quad \omega=0.79, \quad
\chi^2/\mbox{dof} = 1.25, \\
\beta_\rmi{vol} & = & 0.45485(37),\quad \nu=0.77(8), \quad \omega=0.66(19), 
\quad \chi^2/\mbox{dof} = 0.07. \la{svol}
\ea

We observe that 
the location of $\beta_\rmi{vol}$ depends on the 
definition used in constructing macroscopic vortex loops
in a statistically significant way. The 
ambiguity inherent in this procedure is due to discretization 
(lattice) effects, cf.\ \fig\ref{fig:connect}.   
Moreover, for the maximal connectivity definition, 
the percolation critical point is $\beta_\rmi{vol}\sim 0.461(1)$, 
far above the thermodynamic critical point, 
as is already apparent in \fig\ref{fig:betap}.  
Even for the stochastic connectivity definition, $\beta_\rmi{vol}$
does not coincide with the $\beta_c$, with 
a difference on the level of $(2...3) \sigma$ depending slightly
on the ansatz used for the $N\to\infty$ extrapolation. 
For a more precise statement we rely below on \eq\nr{inequ0}
and $\beta_\rmi{mass}$.

\subsection{Closed loop order parameter}

\begin{figure}[tb]

\centerline{
\epsfxsize=6.5cm\epsfbox{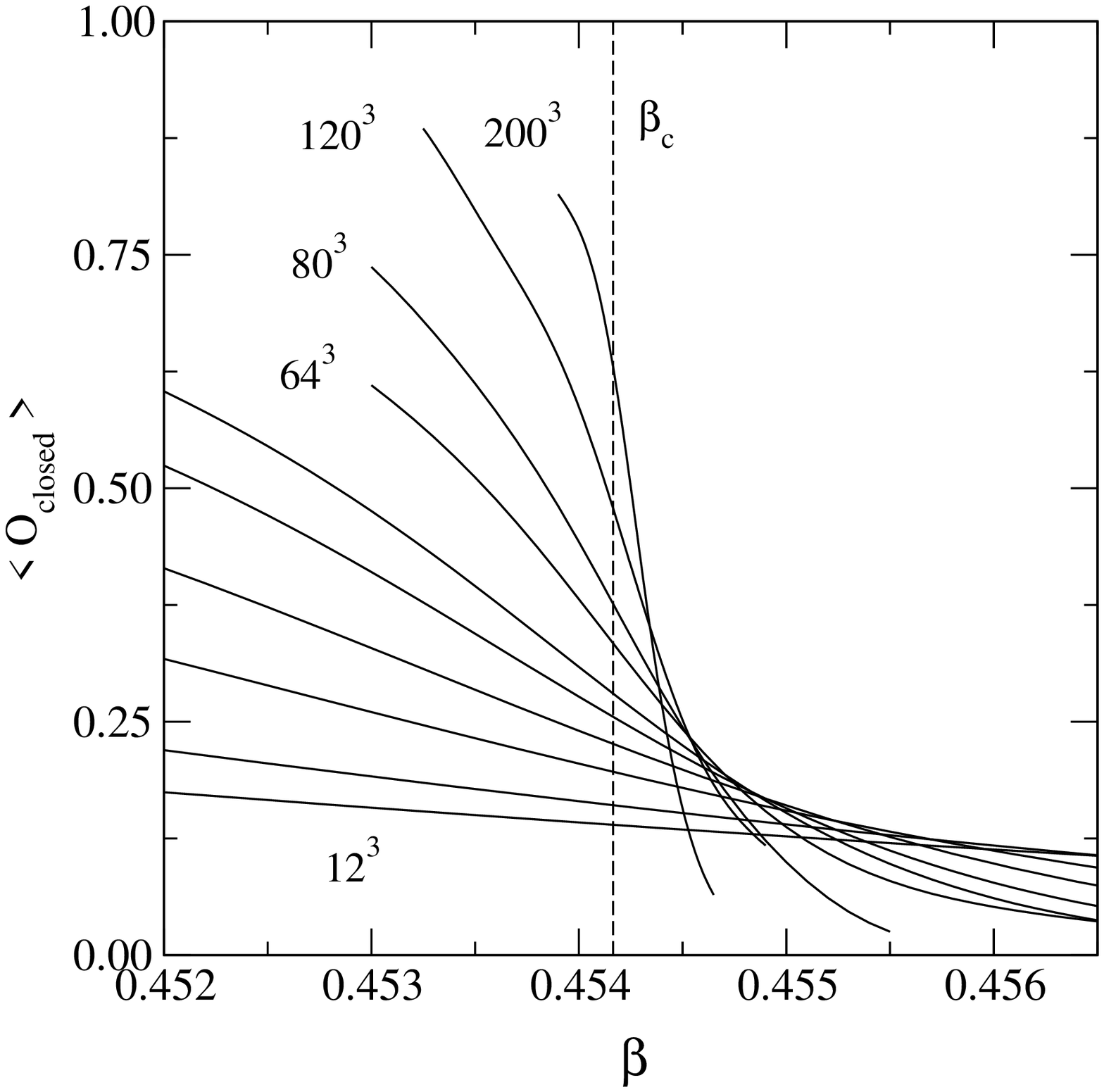}\hspace*{0.5cm}%
\epsfxsize=6.5cm\epsfbox{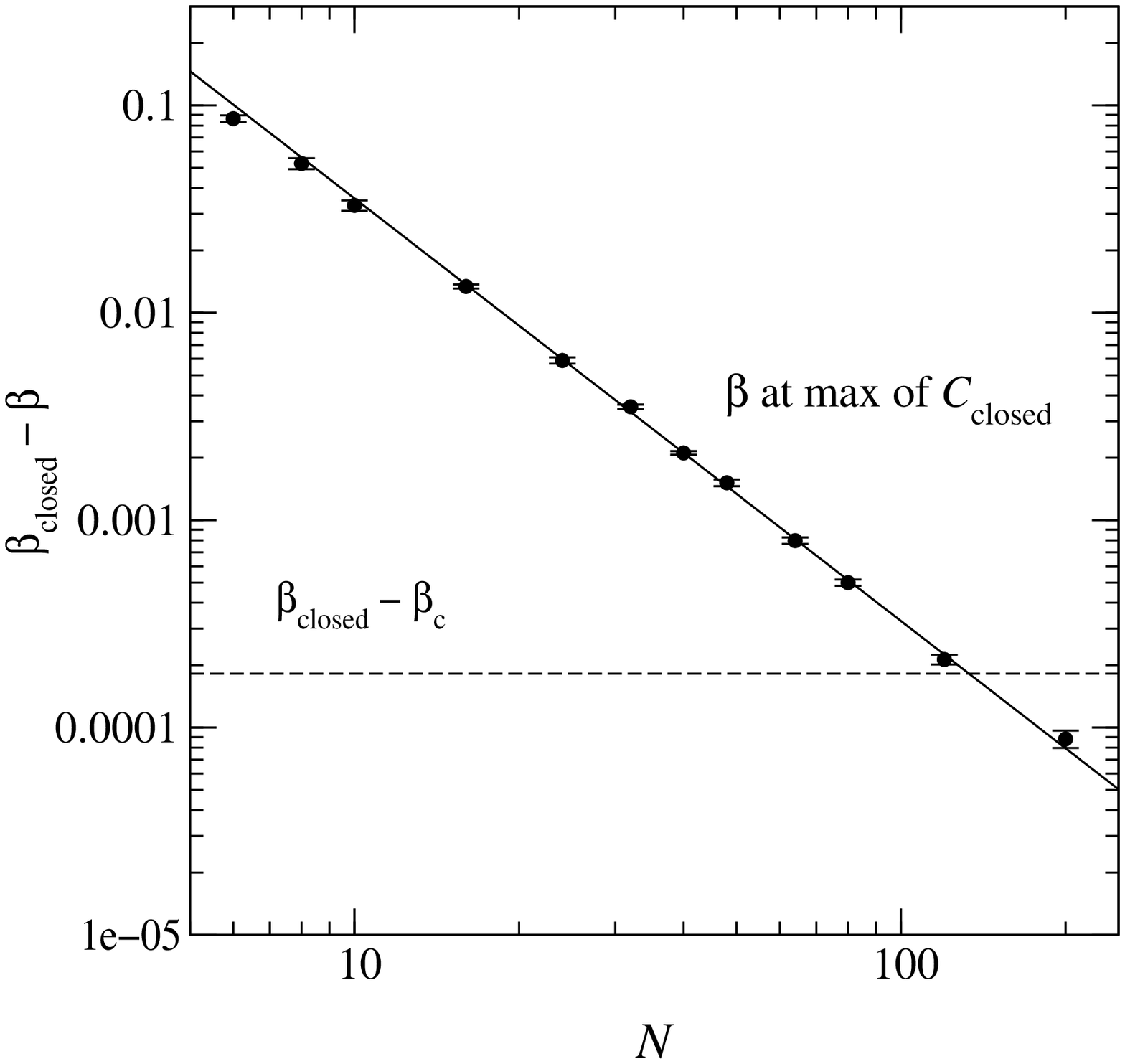}}

\caption[a]{Left: the reweighted  
$\langle O_\rmi{closed}\rangle $ as a function of $\beta$.  
Right: the apparent percolation point as obtained from 
the position of the maximum of $C_\rmi{closed}$
(or, for an observable only having values 0, 1, 
from the point where $\langle O_\rmi{closed}\rangle =0.5$), 
together with the fit in \eq\nr{bpclosed} whereby
$\beta_\rmi{closed}=0.45435$.  The log-log plot shows clearly 
that the power law describes the data well and $\beta_\rmi{closed} > \beta_c$
for $N\to\infty$.}  
\la{fig:closed}
\end{figure}

Consider then $O_\rmi{closed}$.
As conjectured by \eq\nr{inequ}, $\beta_\rmi{closed}$
can be even closer to $\beta_c$ than $\beta_\rmi{vol}$, 
and we have therefore carried out a more precise 
analysis with volumes up to $200^3$.
The behaviour is shown in \fig\ref{fig:closed}. 
A fit according to \eq\nr{fss} to volumes $\ge 16^3$ with 
one free exponent gives
\ba 
\beta_\rmi{closed} & =  & 0.45435(3), \quad \nu=0.490(6), \quad 
\chi^2/\mbox{dof} =9.5/8. \la{bpclosed}
\ea
The data is not accurate enough 
to allow for a good fit with two free exponents.
A fit with exponents fixed to the thermodynamic values of $\nu$
and $\omega$ has $\chi^2/\mbox{dof} =53/9$, and thus a very bad
confidence level $\sim 2\times 10^{-8}$.
 
The central value of $\beta_\rmi{closed}$ 
in \eq\nr{bpclosed} is now much closer to $\beta_c$
than in \eq\nr{svol}. However the errors are also 
an order of magnitude smaller, due to the 
larger volumes used and the smaller range of reweighting in $\beta$.
Thus the discrepancy is $\sim 6 \sigma$. 
The disagreement can be very 
clearly seen in \fig\ref{fig:closed}(right). 

We do not consider separately the case of maximal connectivity,
since it would increase $\beta_\rmi{closed}$ moving it further
away from $\beta_c$. Likewise, \eq\nr{inequ0} shows that
$\beta_{x}$ and $\beta_{xyz}$ cannot coincide with $\beta_c$, either.

\subsection{Mass order parameter}

\begin{figure}[tb]

\vspace*{-0.5cm}

\centerline{
    \psfig{file=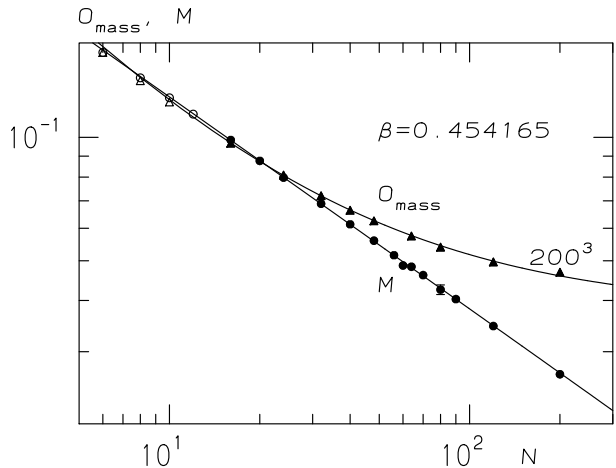,angle=270,width=10cm}}

\vspace*{-0.5cm}

\caption[a]{The behaviour of $\langle O_\rmi{mass}\rangle$
at $\beta_c$, compared
with the thermodynamic quantity $M$ defined in \eq\nr{EM}
(more precisely, 
$M/4$ in order to have similar magnitudes). 
}  
\la{fig:masscrit}
\end{figure}

Since both $\beta_\rmi{vol}$ and $\beta_\rmi{closed}$ are above $\beta_c$,
\eq\nr{inequ} suggests that $\beta_\rmi{mass}$ may be very close 
to it. A direct determination of $\beta_\rmi{mass}$ is difficult because
it turns out that there is no clear peak in $C_\rmi{mass}$ up to 
very large volumes, due to strong fluctuations in the symmetric
phase, and thus it is not easy to obtain
finite volume estimates for $\beta_\rmi{mass}$.
Therefore, we use a different technique to demonstrate that 
$\beta_\rmi{mass}>\beta_c$.

We stay precisely at $\beta_c$, given in \eq\nr{betac}, and measure
$\langle O_\rmi{mass}\rangle$ on lattices of sizes
up to~$200^3$. The results are shown
in \fig\ref{fig:masscrit}.
Recall that we expect a vanishing result at $\beta_\rmi{mass}$, 
and a finite value in the percolated phase $\beta<\beta_\rmi{mass}$.
Since $\langle O_\rmi{mass}\rangle$  behaves similarly to magnetization 
with an inverted $\beta$-axis, we also show the magnetization
in the same figure. 

A fit to volumes $20^3\ldots200^3$ with a single free exponent
gives $0.60/N^{0.61(4)}+0.0393(7)$, with $\chi^2/\mbox{dof} =
7.3/5$.  This clearly shows a non-vanishing value in the thermodynamic
limit. For the magnetization, a fit with a free exponent 
with volumes  $20^3 \ldots 200^3$ gives
$1.1/N^{0.510(9)}-0.002(3)$, $\chi^2/$dof $=0.49$, consistent 
with $\beta/\nu = 0.52$ in \eq\nr{table}
and a vanishing result at $N\to\infty$. 
Thus, while magnetization 
exhibits the expected critical behaviour, from the 
point of view of $O_\rmi{mass}$  we are already in the 
percolated phase, showing that $\beta_\rmi{mass} > \beta_c$.
In the light of \eq\nr{inequ0}, this also confirms
that $\beta_\rmi{vol} > \beta_c$.

\subsection{Line tension}
\la{linetension}

\begin{figure}[tb]

\centerline{
    \psfig{file=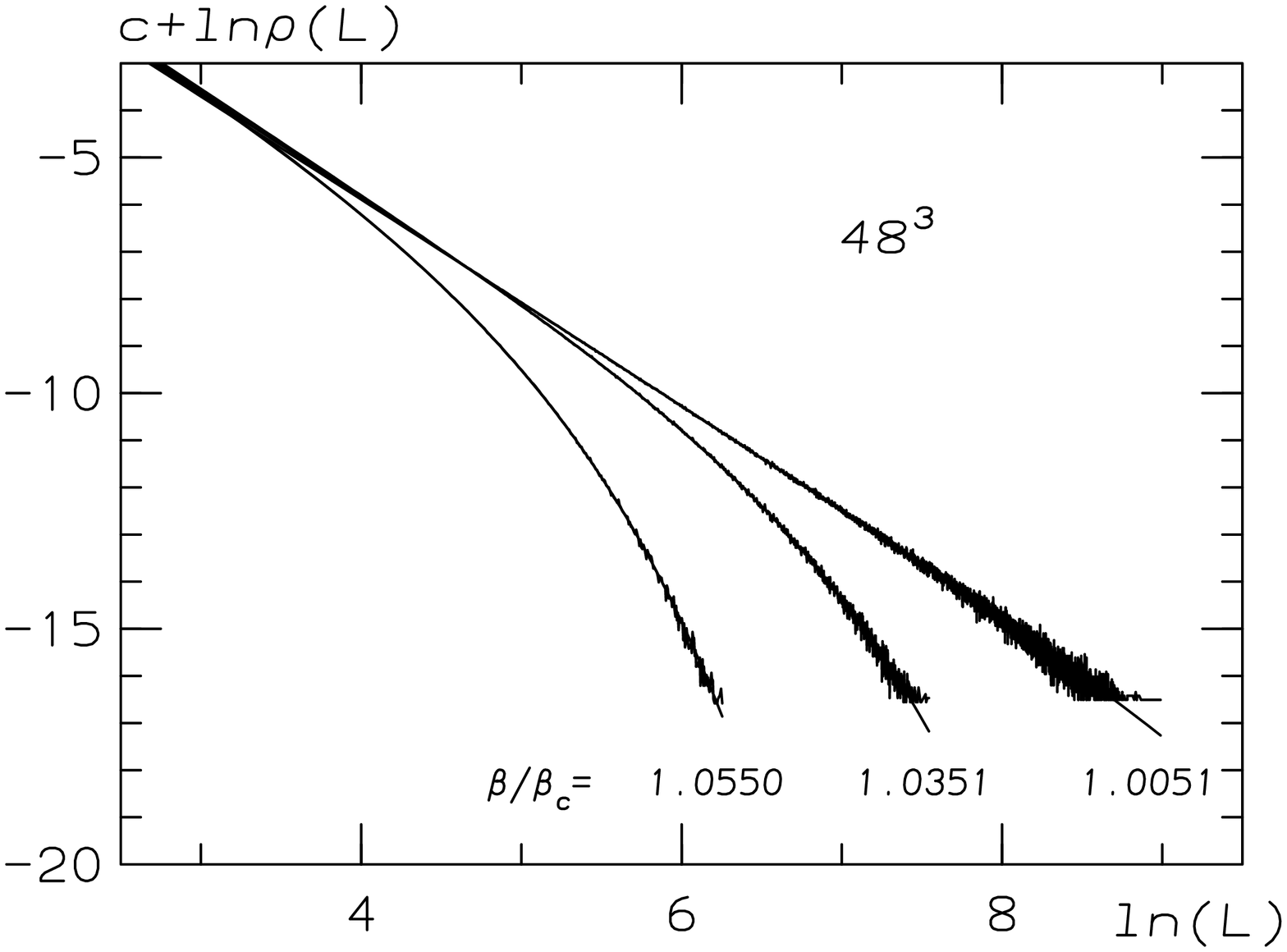,angle=270,width=7.7cm}
    \psfig{file=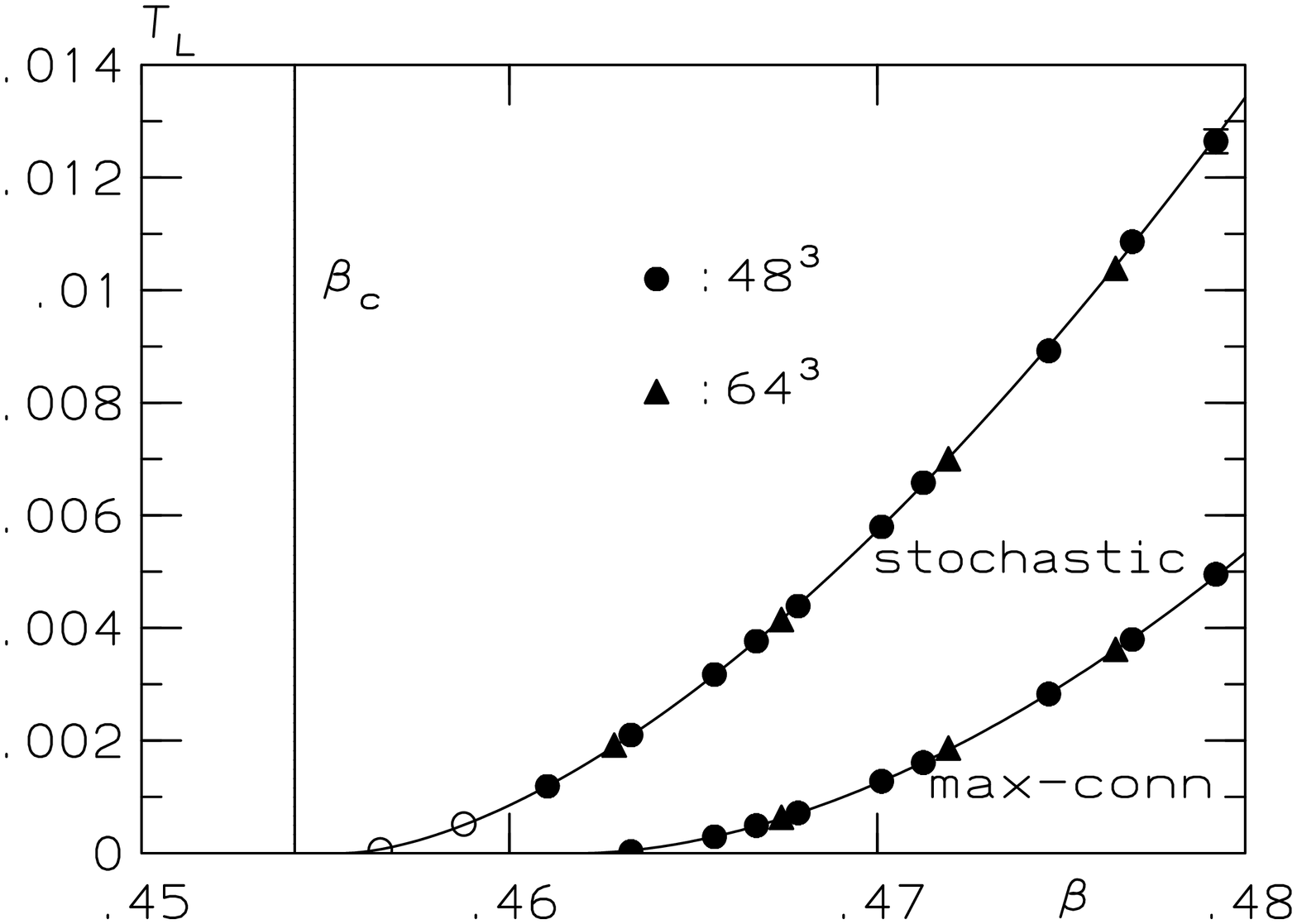,angle=270,width=7.7cm}}

\caption[a]{Left: The length distribution of vortices
at $\beta$ slightly above $\beta_c$ with the stochastic definition. 
Right: the line tension, determined from the 
fit in \eq\nr{loopdist} with a free
parameter $\alpha$, with both definitions.}  \la{fig:Tl}
\end{figure}

Let us finally discuss the line tension.
The data is shown in \fig\ref{fig:Tl}. Since the analysis
of $T_L$ requires simulations at several values of $\beta > \beta_c$,
we have performed the analysis using volumes
$48^3$ and $64^3$ only. We will again monitor 
$\beta_{T_L}$, as well as a critical exponent $\gamma_{T_L}$, defined by
\be
\la{ltdef}
T_L\propto|\beta-\beta_{T_L}|^{\gamma_{T_L}}.
\ee
Since this behaviour assumes the infinite volume case, we
cannot in practice go too close to $\beta_c$ where the finite
volume cuts off the critical fluctuations. At the same time, 
we should be close enough to $\beta_c$ to be in the critical 
regime. In practice, we have chosen the window
$\beta\in(0.46,0.48)$ to satisfy these
requirements, and have checked the volume independence
at a couple of $\beta$, see \fig\ref{fig:Tl}(right). 

Using the stochastic definition, 
a fit to \eq\nr{loopdist}
in this range gives $\alpha=2.11(2)$, and $T_L$ 
as shown in \fig\ref{fig:Tl}(right). Fitting  this $T_L$
to \eq\nr{ltdef}, 
as well as the one obtained with a fixed $\alpha=2.5$,
we get
\ba
\alpha & = & 2.11(2), \quad 
\beta_{T_L}=0.4555(2),\quad \gamma_{T_L}=1.62(3),\quad 
\chi^2/\mbox{dof} = 0.27, \\
\alpha & = & 2.5, \quad 
\beta_{T_L}=0.4558(4), \quad \gamma_{T_L}=1.67(5),\quad \chi^2/\mbox{dof} =2.7.
\ea
If $\beta_{T_L}$ is fixed by hand to a smaller value,
$\gamma_{T_L}$ can go up to $\sim 1.72$. The maximal 
connectivity definition gives clearly contradicting numbers, 
see \fig\ref{fig:Tl}(right), with an exponent $\gamma_{T_L} = 1.80(3)$.

In conclusion, the percolation point $\beta_{T_L}$ again differs
significantly from the critical point in \eq\nr{betac}. Moreover, 
the values we find for $\gamma_{T_L}$ do not coincide with any of the 
thermodynamic exponents.\footnote{Our result for the exponent 
$\gamma_{T_L}$ does not agree with $\gamma_{T_L}=1.45(5)$
measured in~\cite{ns}, nor with $\gamma_{T_L}=1.50(1)$
measured in~\cite{Antunes}. However, the fitted $\beta$ ranges 
there were about 10~\cite{ns} or more~\cite{Antunes} times
wider than in our case, so that corrections to asymptotic 
scaling could play a role.}

\section{Conclusions}

It is quite obvious that in discrete spin models even naive geometrically 
defined percolation observables show structure close to the 
thermodynamic phase transition. Indeed, the ordered and disordered
phases of the spin models are quite well understood, and either 
almost frozen (in the ordered phase), or with almost random
fluctuations (in the disordered phase), which guarantees that 
percolation seems to take place. The question we have 
addressed here is whether this rough agreement can be promoted
to a precise level in the vicinity of the thermodynamic 
transition point, where the spin models show
universal critical behaviour, and allow to 
define a non-trivial continuum field theory. 
 
We have found that in the three-dimensional XY model, 
the geometrically defined
percolation transition of the vortex loops is extremely 
close to the thermodynamic phase transition, but does not 
coincide with it with any of the 
observables we have used. There is a clear discrepancy between
the ``maximal'' and ``stochastic'' definitions used for constructing 
macroscopic vortex loops. Even with the stochastic definition, a 
finite-size scaling analysis at the thermodynamic critical point,
whose location is known very accurately from previous studies, 
shows the absence of any true critical behaviour. 
We of course cannot exclude that it would be possible to 
bring the percolation transition 
even closer to the thermodynamic one by very carefully fine-tuning the 
percolation criteria;  however, this kind of a procedure 
would have no predictive power.

We believe that these conclusions apply also to more realistic 
theories, such as the Ginzburg-Landau theory of superconductivity
and non-Abelian gauge theories. 
These theories are typically much more complicated
than the XY model used here, and therefore it is in most cases
very difficult to achieve the level of accuracy needed to
confirm this in a Monte Carlo simulation. However, we may note
that the observation of a percolation transition related to 
so called $Z$-vortices in crossover regime of the electroweak 
theory~\cite{cgis2} bears a similarity with our conclusion.

Finally, we would like to emphasise that
although geometrically defined percolation 
observables may not reflect the properties
of the true phase transition in a rigorous sense, 
the description of the system in terms of the vortex loop degrees
of freedom can still be useful, if a good observable is chosen. 
This is certainly true for the tension~$T$
as discussed in the Introduction, and 
might also be the case in the non-equilibrium 
dynamics well after a cosmological phase transition,
when the vortex loops cease to be merely thermal fluctuations and become
macroscopic classical objects~\cite{reviews}. 


\section*{Acknowledgements}

The simulations were carried out with a number of workstations
at the Helsinki Institute of Physics and at Nordita, as well as 
with an Origin 2000 at the Center 
for Scientific Computing, Finland.
This work was partly supported by the 
TMR network {\em Finite Temperature Phase
Transitions in Particle Physics}, EU contract no.\ FMRX-CT97-0122.
T.N. acknowledges support from DFG and discussion with H. Kleinert, 
and A.R. was partly supported by the University of Helsinki.


\appendix
\renewcommand{\theequation}{\Alph{section}.\arabic{equation}}


\end{document}